\documentclass[conference]{IEEEtran}

\IEEEoverridecommandlockouts
\usepackage{cite}
\usepackage{todonotes}
\usepackage{amsmath,amssymb,amsfonts}
\usepackage{algorithmic}
\usepackage{graphicx}
\usepackage{textcomp}
\usepackage{xcolor}
\usepackage{booktabs}
\usepackage{glossaries}
\def\BibTeX{{\rm B\kern-.05em{\sc i\kern-.025em b}\kern-.08em
    T\kern-.1667em\lower.7ex\hbox{E}\kern-.125emX}}
\IEEEoverridecommandlockouts \IEEEpubid{\makebox[\columnwidth]{979-8-3315-6128-4/26/\$31.00~\copyright{}2026 IEEE \hfill} \hspace{\columnsep}\makebox[\columnwidth]{ }} 
\makeglossaries

\newacronym{XR}{XR}{Extended Reality}
\newacronym{ISAC}{ISAC}{Integrated Sensing and Communications}
\newacronym{DL}{DL}{Deep Learning}
\newacronym{CNN}{CNN}{Convolutional Neural Network}
\newacronym{mmWave}{mmWave}{Millimeter-Wave}
\newacronym{COTS}{COTS}{Commercial Off-The-Shelf}
\newacronym{PPBP}{PPBP}{Power-per-beam-pair}
\newacronym{pp}{pp}{percentage points}
\newacronym{SNR}{SNR}{Signal-to-Noise Ratio}
\newacronym{LOUO}{LOUO}{Leave-one-user-out}
\newacronym{CSI}{CSI}{Channel state information}
\newacronym{OFDM}{OFDM}{Orthogonal Frequency-Division Multiplexing}
\newacronym{LoS}{LoS}{Line-of-sight}
\newacronym{sEMG}{sEMG}{surface electromyography}
\newacronym{SSB}{SSB}{Synchronization signal block}

\begin{document}

\title{Integrated Sensing and Communications for Real-time Avatar Control in XR over 5G}
\author{
\IEEEauthorblockN{
Nabeel Nisar Bhat\IEEEauthorrefmark{1},
Javad Sameri\IEEEauthorrefmark{2}\IEEEauthorrefmark{4},
Rreze Halili\IEEEauthorrefmark{1},
Rafael Berkvens\IEEEauthorrefmark{1},
Maria Torres Vega\IEEEauthorrefmark{4},
and Jeroen Famaey\IEEEauthorrefmark{1}
}

\IEEEauthorblockA{\IEEEauthorrefmark{1}IDLab, University of Antwerp - imec, Antwerp, Belgium}

\IEEEauthorblockA{\IEEEauthorrefmark{2}IDLab, Ghent University - imec, Ghent, Belgium}

\IEEEauthorblockA{\IEEEauthorrefmark{4}Light and Lighting Technology Lab, KU Leuven, Ghent, Belgium}
}
\maketitle

\begin{abstract}
Extended Reality (XR) presents a challenging use case for 5G and 6G networks, requiring high data-rates and low-latency communication to deliver a truly immersive experience. Moreover, in order to seamlessly translate physical actions to the virtual world, accurate gesture recognition and pose estimation are required. Current XR interaction solutions based on handheld controllers and cameras cannot easily capture full-body poses, inhibit the free use of hands, and require good visibility and a clear line of sight. In this work, we propose a multi-modal sensing architecture for XR that combines 5G Millimeter-Wave (mmWave) Integrated sensing and communication (ISAC) and surface electromyography (sEMG) signals. 5G mmWave ISAC cannot only be used to deliver content wirelessly to the Head-mounted display (HMD), but also the same communication signals can be used to derive coarse body-level gestures and poses of the user, to support real-time avatar control. For fine-grained finger-level gestures, our architecture leverages lightweight sEMG sensors that capture forearm muscle activity. To illustrate the need of both modalities, we present evaluations of both sensing technologies. At the body level (5G), our architecture relies on power-per-beam-pair (PPBP), which can be computed from standard beam management or beam sweeping procedures of the 5G NR standard. PPBP-based sensing achieves 82.2$\pm$5.9\% average accuracy when evaluated on users not seen during training. 
For fine-grained finger-level interactions, we show that \gls{sEMG} carries strong discriminative information achieving consistent promising performance across different movement settings. Thus, combining the two modalities enables multi-scale gesture recognition, at the body level via existing 5G signals and finger level via lightweight sEMG sensors, forming a complete XR framework.

\end{abstract}

\begin{IEEEkeywords}
eXtended Reality, 5G, surface electromyography, Human Pose Estimation 
\end{IEEEkeywords}

\section{Introduction}
\gls{XR}, encompassing augmented reality (AR), virtual reality (VR), and mixed reality (MR), is considered a key use case for 5G and 6G networks \cite{alhakamy2024extended, ahmad2023leveraging}. These applications pose stringent requirements such as very high data-rates (hundreds of Mbps to several Gbps for 4K/8K) and very low motion-to-photon latency ($<20$ milliseconds) \cite{10634005} to enable truly immersive experiences. Further, accurate pose estimation and gesture recognition capabilities are desired to enable seamless interactions. Current \gls{XR} systems rely on handheld controllers, RGB cameras, and depth sensors to translate physical actions into the virtual world. However, these methods cannot easily capture full-body poses, inhibit free use of hands, and require good visibility and a clear \gls{LoS}. 

5G \gls{mmWave} frequency bands (24.25 - 71.0 GHz) \cite{11161113}, offer promising data rates (with typical peaks up to 20 Gbps), due to the wide bandwidth and low-latency. Additionally, shorter wavelengths at these frequencies enable compact large-scale antenna arrays, which, together with wide bandwidth, improve range and angular resolution \cite{9898900,9330512} for sensing applications such as localization, and gesture recognition. Thus, the 5G NR standard provides a powerful sensing platform that can enable device-free interactivity in XR applications, as a \textit{by-product} of existing communication infrastructure. This is referred to as \gls{ISAC} \cite{11112522}. The major advantage of this approach is that sensing can be performed at a limited additional cost. 

However, despite these advantages, the spatial resolution of 5G mmWave sensing remains insufficient to reliably capture fine-grained hand and finger articulations that are essential for natural interaction in XR~\cite{MultisenseVR2026}. Prior work has explored the use of \gls{sEMG} signals to capture muscle activity and infer hand gestures, enabling interaction even when the hands are out of sight or occluded \cite{sivakumar2024emg2qwerty}.

Herein, we envision 5G mmWave ISAC to be used for coarse body-level interactions and poses, while lightweight sEMG sensors can capture finger-level interactions, thus providing a complete XR interaction pipeline. In this paper, we present a multi-modal sensing system architecture, sensing feature selection, communication-sensing trade-off, and complementary sensing modalities, along with an in-depth experimental evaluation of gesture recognition in a cross-user setting. In addition, we provide a brief overview of XR-related use cases in the Third Generation Partnership Project (3GPP) and the European Telecommunications Standards Institute (ETSI) that can be enabled and enhanced by ISAC capabilities.

\begin{figure}[t]
    \centering
    \includegraphics[width=\columnwidth,trim=1cm 0cm 0cm 1cm, clip]{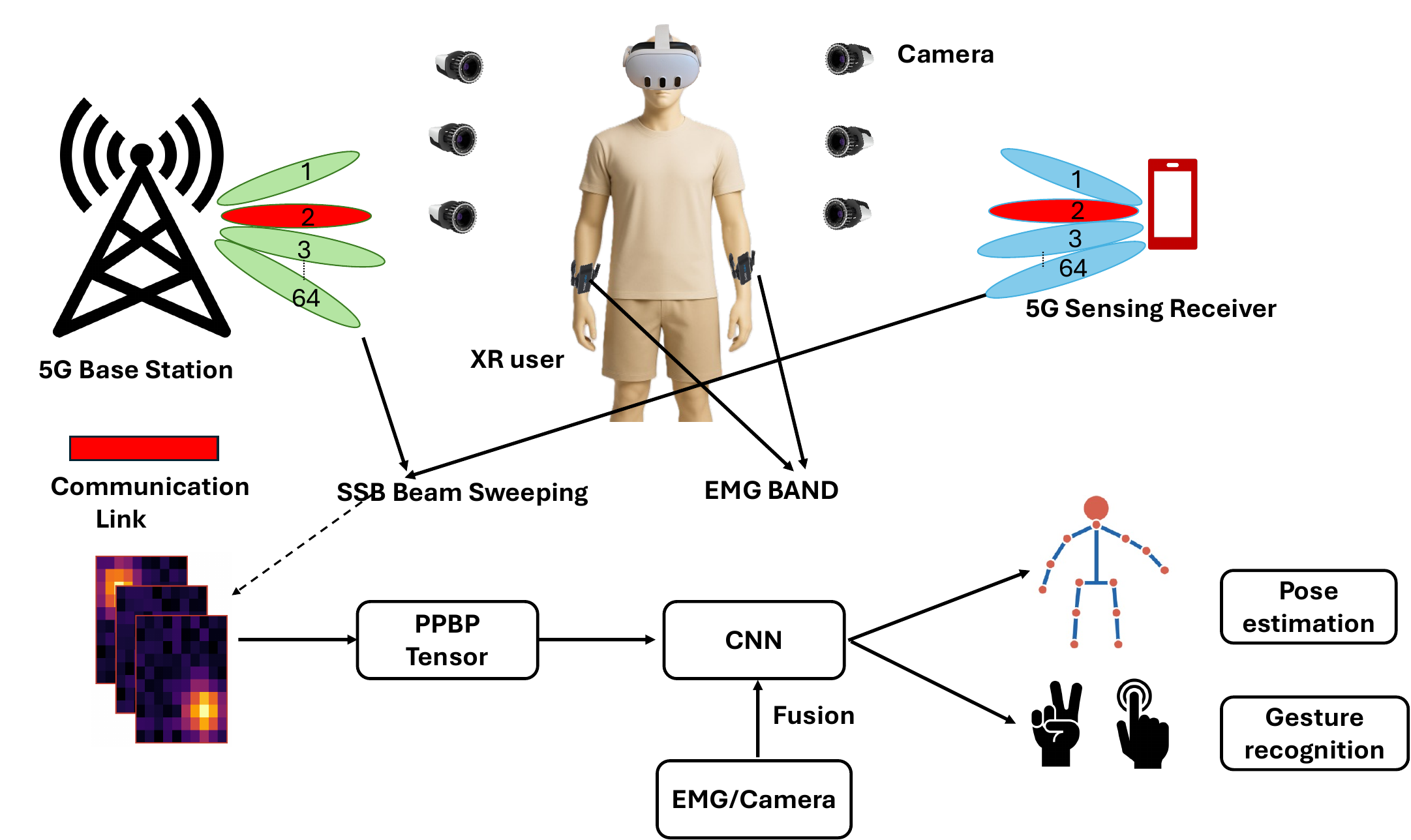}
    \caption{5G ISAC system architecture for immersive XR: SSB beam sweeping used to derive PPBP, which is then fed to a CNN for a specific downstream task. The camera is used only for training to regress poses, with EMG providing finger-level tracking.}
    \label{fig:system_arhcitecture}
\end{figure}

\section{5G ISAC and sEMG for Human Pose Estimation}

Figure \ref{fig:system_arhcitecture} shows the 5G ISAC and sEMG architecture for immersive XR. The system consists of a gNodeB base station that acts as a transmitter and a sensing receiver in a bi-static configuration, to allow passive sensing of users in the coverage area.
The gNodeB transmits a standard 5G NR mmWave OFDM waveform in the downlink, including periodic \gls{SSB} \textit{beam sweeping} \cite{10797375} to help UEs identify the best beam for communication. At mmWave frequencies, up to 64 directional beams are transmitted in sequence over a 5-millisecond window, using directional antennas. UEs measure the reference signal received power (RSRP) per SSB beam and report the best beam index to the gNodeB. Then the communication proceeds through the best beam. This procedure is repeated every 20 milliseconds.
While the gNodeB performs SSB beam sweeping, the sensing receiver can compute \gls{PPBP}, the received power per transmit-receive beam. A user in the environment modulates the power across the spatial grid, which can be tracked over time to infer the user's pose and gestures. Thus, a temporal window of \gls{PPBP} is fed to the deep learning model, such as a convolutional neural network (CNN), for a specific downstream task. In this paper, we consider gesture recognition as a downstream task, but in our previous work \cite{11112522}, we showed that PPBP can be used for pose estimation as well.

While 5G mmWave sensing can accurately capture coarse body movements and spatial dynamics, its spatial resolution remains insufficient to reliably distinguish fine-grained finger articulations required for natural XR interaction. Subtle hand gestures such as pinching, grasping, or finger flexion produce limited variations in reflected radio signals, making them difficult to infer using mmWave sensing alone. To address this limitation, we envision \gls{sEMG} as a complementary sensing modality that can directly measure muscle activation patterns in the forearm corresponding to finger and hand movements. Unlike vision-based approaches, \gls{sEMG} does not require \gls{LoS} and remains robust under occlusion or challenging lighting conditions. By combining mmWave sensing with \gls{sEMG}, the system can jointly capture both global body motion and fine-grained interaction cues. 

\subsection{Sensing Signal Representations in 5G ISAC}

\textbf{\gls{CSI}}: In ISAC, channel state information (CSI), which characterizes how the signals propagate from transmitter to receiver over a wireless channel, is widely used as a fine-grained sensing feature for various downstream applications \cite{chen2023cross}. The CSI captures both spatial and temporal variations of the wireless channel. However, in bi-static ISAC systems with physically separated transmitter and receiver, each having its own local oscillator, the CSI phase is affected by various hardware imperfections such as carrier frequency offset (CFO), and time-varying phase noise. CSI
is heavily affected by environment propagation paths and user morphology, making it challenging to generalize to unseen users. Further, the distribution gap between CSI measurements corresponding to different users performing gestures or activities is quite significant. As a result, conventional domain adaptation techniques, such as Adversarial Discriminative Domain Adaptation (ADDA), which perform well on vision tasks, do not translate well to CSI-based sensing. Moreover, CSI is not exposed in commercial-off-the-shelf devices without any firmware modifications, thereby introducing additional overhead. Finally, the high dimensionality of CSI (subcarriers × antennas × time) increases computational complexity when using it as a feature for deep learning.

\textbf{Micro-doppler:} Phase of CSI can be tracked over time, to compute micro-doppler, which captures a fine-grained velocity profile of moving body parts. Micro-doppler features are domain-independent \cite{pegoraro2025disc}, since they capture only motion dynamics, hence are less affected by environment-specific multi-path or user-specific body morphology.
However, accurate micro-doppler computation requires precise calibration of the phase, along with fixed inter-frame spacing, which is not always feasible or relies on sparse recovery methods which are computationally expensive \cite{pegoraro2022sparcs}. 

\textbf{Power-per-beam-pair (PPBP)}: 5G NR inherently provides beam-level power measurements as part of standard operation, 3GPP Release 15 onwards. 5G gNodeB performs periodic SSB beam sweeping to help UEs identify the best beam for communication \cite{10797375}.
Compared to CSI and micro-doppler, PPBP is simple to compute, does not require any \textit{phase calibration} and can be extracted without any modification to the \textit{5G protocol stack}. These properties make PPBP a practical and scalable sensing feature for 5G ISAC applications. As such, PPBP will be used in the remainder of this paper.

\subsection{\gls{sEMG} Signal processing for finger-level interactions}

\gls{sEMG} provides a natural complement to 5G ISAC for immersive XR by capturing the electrical activity of forearm muscles during hand and finger motion, including interaction cues that are difficult to infer from mmWave sensing alone. In practice, these signals can be acquired using lightweight wearable armbands placed around the forearm; in our multi-modal setup, two 8-channel armbands record muscle activity at 500~Hz \cite{MultisenseVR2026}. Since raw \gls{sEMG} is susceptible to motion artifacts and electrical interference, it must first undergo pre-processing, after which relevant feature descriptors can be extracted to characterize the amplitude, variability, and spectral distribution of muscle activation patterns associated with XR interactions. These features can then be used as input to lightweight classifiers to recognize gesture primitives relevant to immersive applications, such as pinch and fist-grab, which are central to object manipulation in XR~\cite{MultisenseVR2026}.

\subsection{Communication and Sensing trade-off}
5G mmWave systems provide a wide channel bandwidth up to 800 MHz via carrier aggregation. This provides peak data rates of 20 Gbps \cite{11161113} and latency less than 1 millisecond under LoS conditions \cite{fezeu2023depth,shah2022comparative}. These characteristics make 5G mmWave systems particularly attractive for XR applications, where high data-rates and low-latency are critical. However, 5G ISAC systems need to divide airtime between communication and sensing. The impact on communication is straightforward; the throughput is proportional to the communication airtime. On the other hand, the impact on sensing is not straightforward. This trade-off was recently quantified in our previous work \cite{11366454} using the same testbed and dataset for gesture recognition. Our results indicated that allocating 25\% of the airtime to sensing reduces gesture classification accuracy by only 0.15 \gls{pp}, compared to full-time sensing.
\section{XR-related ISAC use cases in Standardization} 
In 3GPP 5G and 6G systems, ISAC refers to the capability of using the same wireless radio signals and network infrastructure for both communication and sensing \cite{3gpp_TR22837}.
The latest version of 3GPP TR 22.837 (Release 19) describes a set of use cases and potential requirements to improve the 5G system to support integrated sensing services \cite{3gpp_TR22837}. 

The \textit{Seamless XR streaming} use case shows how sensing information collected from both 3GPP sensors in the RAN and UE (reflected, refracted, and diffracted RF signals) and non-3GPP sensors (IMUs, LiDAR cameras, and position sensors) can be fused to construct a Sensing RF Map of the surrounding environment. This enables the network to consider channel changes and maintain reliable wireless links for XR services.

The \textit{Immersive experience based on the sensing} use case shows how RF signal reflections such as Doppler shift, time-of-flight, and amplitude variations can detect and track users in an environment without requiring them to carry a device. The sensing results can then be used by smart devices (e.g., screens, speakers, and lights) to dynamically adapt audio and visual effects and create context-aware immersive experiences. 

The \textit{Coarse gesture recognition for application navigation and immersive interaction} use case demonstrates how NR-based RF sensing can detect human gestures and body movements for touchless control and XR interaction. In this scenario, range and Doppler are extracted from 5G sensing signals and combined with non-3GPP sensor data to improve recognition accuracy and enable applications such as avatar control, hand tracking, and touchless device navigation.

These use cases highlight multi-modal sensing (RF, visual, and motion sensing), cooperative sensing between UE and RAN, and distributed architectures where sensing data can be processed at the device, edge, or 5G core and shared with immersive applications. Details of the service requirements for ISAC based on 5G are found in 3GPP TS 22.137 \cite{3gpp_TS22137}, while the system and RAN architectural framework for 6G ISAC are found in ETSI GR ISC 003 \cite{etsi_GR_ISC003}. 
3GPP TR 22.870 (Release 20) outlines high-level principles of some XR-related use cases to identify potential requirements for enabling the 6G system.

The \textit{Enhanced XR user navigation, Gesture recognition in industrial environments, and Structural health monitoring} use cases highlight the need for higher sensing resolution, larger sensing coverage, lower latency, and the integration of AI/ML-based sensing techniques to support context-aware services \cite{3gpp_TR22870}. In addition, the report includes a dedicated chapter on immersive communication where ISAC is introduced as a supporting 6G capability.

This paper contributes to the use case on \textit{Coarse gesture recognition for application navigation and immersive interaction}, by presenting a multimodal approach that combines 5G mmWave ISAC with sEMG data for coarse-level gesture recognition combined with accurate finger tracking.

\section{Validation and Results}
To illustrate the need of both modalities, this Section first presents evaluations for body level 5G body pose recognition (Section~\ref{sec:sec:5GISACev}). It then, introduces evaluation of sEMG signals for fine-grained pose analysis (Section~\ref{sec:sec:sEMGev}).

\subsection{Experimental evaluation of 5G ISAC for body tracking}\label{sec:sec:5GISACev}

For the 5G mmWave ISAC validation, we used our 5G mmWave ISAC testbed~\cite{bhat2025mmgan,11366454} consisting of a bi-static setup with two Sivers EVK's (EVK 06002), functioning as transmitter and receiver. The system cycles through $50\times56=2800$ beam pairs at 154 sweeps per second, each beam pair active for approximately 2.3 microseconds. 5G OFDM waveform is used with 760 MHz bandwidth at 60 GHz carrier frequency \cite{bhat2025mmgan}. For the dataset collection, 8 users performed 8 gestures in 7 independent 10-second sessions. The gestures include static poses such as Empty (E) - standing still (serves as baseline for other gestures), Left Lean (LL) - leaning to the left, Right Lean (RL) - leaning to right and dynamic gestures as Push Pull (PP), Left Hand Open (LHO) and Right  Hand Open (RHO) - forward-backward arm movement, where respective hands are extended outward, Arms Up (AU) - raising both arms up, and Arms Open (AO) - both arms extended horizontally to the sides.

As an algorithm, in this work we use a CNN model, \textit{LiteCNN}, consisting of a feature extractor with three convolutional blocks: the first applies \texttt{Conv2d}$(T \to 32,\ 3{\times}3)$ $-$ \texttt{BatchNorm2d} $-$ \texttt{ReLU}; the second    
 applies \texttt{Conv2d}$(32 \to 64)$ $-$ \texttt{BatchNorm2d} $-$ \texttt{ReLU} $-$ \texttt{MaxPool2d}$(2{\times}2)$; and the third applies \texttt{Conv2d}$(64 \to 128)$ $-$ 
 \texttt{BatchNorm2d} $-$ \texttt{ReLU} $-$ \texttt{MaxPool2d}$(2{\times}2)$ $-$ \texttt{Dropout2d}$(p{=}0.3)$. This is followed by \texttt{Linear}$(21504 \to 256)$ $-$
\texttt{ReLU} $-$ \texttt{Dropout}$(p{=}0.5)$ $-$ \texttt{Linear}
$(256 \to 8)$ layer, totalling \textit{5.6\,M parameters}. Moreover, for training we use the Adam optimiser with a learning rate of $10^{-3}$ for 30 epochs.

Our evaluation follows a \gls{LOUO} setting, where the neural network is trained on 7 users, and tested on an unseen user for cross-domain generalization. Further, for training subjects, 6 sessions, 0-5 are considered in training, while the $7^{\text{th}}$ is kept for validation. 

\begin{figure}[t]
    \centering
    \includegraphics[width=\linewidth]{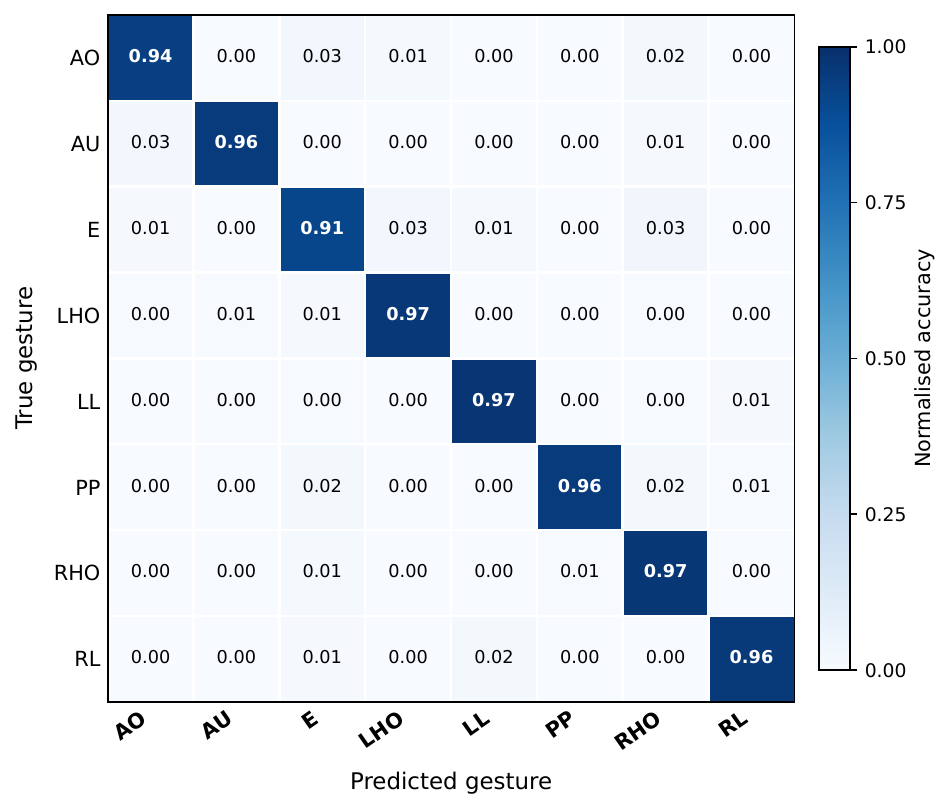}
    \caption{Confusion matrix for the in-domain setting averaged across 8 users.}
    \label{fig:temporal_confusion_matrix}
\end{figure}

First, we establish an in-domain upper bound by training on sessions 0-5 of all users, and reserving session 6 of each user exclusively for testing. This prevents data leakage. Figure~\ref{fig:temporal_confusion_matrix} shows the confusion matrix for the in-domain setting across the 8 users. The model achieves an overall accuracy of $95.4 \pm 0.1$\,\%  with per-gesture accuracy higher than $90\%$. The results are averaged across three random seeds. E, which represents an empty or standing still pose, seems the hardest class. Thus, 
PPBP provides reliable discrimination across all 8 gestures. However, the downside is that the model requires training data from each user, which limits scalability and practical deployment.

Second, the model is tested on the unseen user whose data is not included in the training. For each target user, the model is trained on PPBP data from the remaining seven users. Table \ref{tab:zero_shot_accuracy} shows the zero-shot accuracy on the test data, with no labeled or unlabeled data from the unseen user used during training. For example, for User 1, the training set consists of data from User 2 to User 8. We can see a considerable variation in gesture recognition accuracy across users. For instance,  for User 5 and User 6, the accuracy is $90\%$, while for User 1, 8, and 4, the performance drops significantly. Across 8 users, the average cross-user accuracy is $82.2\pm5.9\%$. A 16.4 \gls{pp} spread in performance across users highlights the generalization issue, indicating a potential risk for real-world deployment.

To further identify the performance issues in the LOUO setting, we break down the classification accuracy into per-gesture metrics. Table \ref{tab:gesture_zero_shot} shows the zero-shot class-wise accuracy results, $\mu$ represents mean, and $\sigma$ represents standard deviation across 8 users. We see that static poses such as RL (Right Lean), LL (Left Lean) achieve very good average accuracy ($>93\%$), with lower $\sigma$ values, suggesting that PPBP pattern variation is consistent across users and independent of user height and body morphology. On the other hand, the Empty (E) pose, wherein the user stands still, achieves the lowest accuracy and highest standard deviation, suggesting that this pose is largely dependent on the user's body morphology. The same result holds for the in-domain setting as well (cf., Figure \ref{fig:temporal_confusion_matrix}). The model shows limited capability in compensating for the idle body's signature, such as height, posture, and standing position. However, this also implies that PPBP contains discriminative features that can be leveraged for person identification. Dynamic gestures achieve intermediate performance. The performance difference is likely due to the beam pairs that are affected by the difference in arm lengths and execution speed. 

\begin{table}[!t]
\centering
\caption{Per-subject zero-shot accuracy: Cross-domain evaluation.}
\label{tab:zero_shot_accuracy}
\begin{tabular}{lc}
\toprule
\textbf{Subject} & \textbf{0-shot Accuracy} \\
\midrule
User 5 & 90.6\% \\
User 6 & 90.4\% \\
User 3 & 83.4\% \\
User 2 & 83.3\% \\
User 7 & 83.0\% \\
User 4 & 76.7\% \\
User 8 & 75.9\% \\
User 1 & 74.2\% \\
\midrule
\textbf{Mean} & \textbf{82.2 $\pm$ 5.9\%} \\
\bottomrule
\end{tabular}
\end{table}
\begin{table}[t]
\centering
\caption{Per-gesture zero-shot accuracy, averaged across 8 users.}
\label{tab:gesture_zero_shot}
\begin{tabular}{lc}
\toprule
\textbf{Gesture} & \textbf{0-shot Accuracy ($\mu$ $\pm$ $\sigma$)} \\
\midrule
RL  & $94.6 \pm 5.0\%$ \\
LL  & $93.2 \pm 4.7\%$ \\
AU  & $89.1 \pm 6.5\%$ \\
PP  & $84.0 \pm 23.4\%$ \\
LHO & $82.1 \pm 14.1\%$ \\
RHO & $78.8 \pm 13.9\%$ \\
AO  & $77.1 \pm 14.0\%$ \\
E   & $62.0 \pm 22.4\%$ \\
\bottomrule
\end{tabular}
\end{table}

\begin{figure}[!t]
    \centering
    \includegraphics[width=\linewidth]{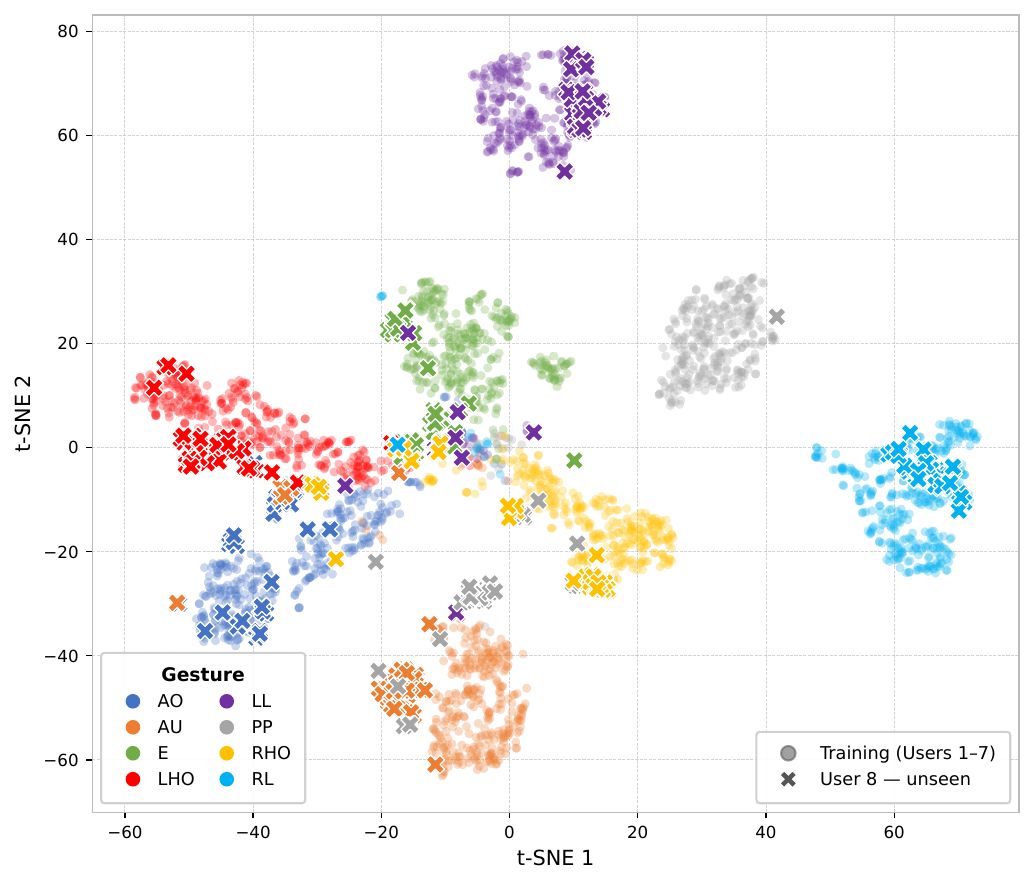}
    \caption{t-SNE embeddings comparison between training users and the unseen user.}
    \label{fig:tsneembeddings}
\end{figure}

Third, Figure \ref{fig:tsneembeddings} shows the domain gap between training users (1-7) and the unseen user (User 8). In particular, t-SNE embeddings \cite{kang2021conditional} are extracted after the ReLU (Rectifier Linear Unit) of the first linear layer in the classification head (before the final $256 \to 8$ projection). Embeddings are colored as per the gesture class, with embeddings for training users shown as circles, whereas for the unseen user, as crosses (X). We can conclude two things from the Figure. First, the backbone or feature extractor has learned well-separated gesture clusters; 8 distinct clusters are clearly visible in the Figure. Second, the embeddings corresponding to the unseen user within each cluster are displaced from the training users. The embeddings of the unseen user cluster are in a sub-region offset from the centroid of the training users. This displacement is likely due to differences in the unseen user's body morphology.

Finally, to further improve the zero-shot performance, instead of feeding raw PPBP to the model, we feed frame-to-frame temporal differences. This is done by first subtracting the mean (computed across $T=20$ time frames) from each frame, to remove the DC offset (static component), and then taking the differences between adjacent time frames.
\begin{equation}
\tilde{P}(t) = P(t) - \frac{1}{T} \sum_{\tau=0}^{T-1} P(\tau), 
\quad t = 0, \ldots, T-1
\end{equation}

\begin{equation}
D(t) = \tilde{P}(t+1) - \tilde{P}(t), 
\quad t = 0, \ldots, T-2
\end{equation}
where $P \in \mathbb{R}^{20 \times 50 \times 56}$. This results in a 19-channel input. The mean subtraction suppresses user-specific static components, while temporal differences ensure only the motion signal is encoded. The result is a feature, \textit{framediff}, that is more user-invariant compared to raw PPBP.
\begin{table}[t]
\centering
\caption{Per-gesture zero-shot accuracy: raw PPBP vs. temporal framediff (mean across 8 held-out users, 3 seeds).}
\label{tab:per_gesture_accuracy}
\begin{tabular}{lccc}
\toprule
\textbf{Gesture} & \textbf{raw PPBP} & \textbf{framediff} & \textbf{$\Delta$} \\
\midrule
RL        & 94.6\% & 93.4\% & $-1.2$ pp \\
LL        & 93.2\% & 92.2\% & $-1.0$ pp \\
AU        & 89.1\% & 90.7\% & $+1.6$ pp \\
PP        & 84.0\% & 88.1\% & $+4.1$ pp \\
LHO       & 82.1\% & 82.8\% & $+0.7$ pp \\
RHO       & 78.8\% & 83.8\% & $+5.0$ pp \\
AO        & 77.1\% & 75.9\% & $-1.2$ pp \\
E & 62.0\% & 74.5\% & $+12.5$ pp \\
\midrule
\textbf{Mean} & \textbf{82.2\%} & \textbf{85.2\%} &  \textbf{$\mathbf{+3.0}$ pp} \\
\bottomrule
\end{tabular}
\end{table}
Table \ref{tab:per_gesture_accuracy} shows the comparison between raw PPBP and framediff in terms of zero-shot accuracy. The overall gain with framediff is 3 \gls{pp}. Class E sees the biggest improvement, as DC offset removal suppresses the user-specific resting pose signature. Other dynamic gestures also benefit, such as PP and RHO. Some gestures, such as RL and LL, show a slight drop in accuracy; however, the differences are minimal. This confirms that features in frame differences are more user-invariant.

\begin{figure}[t]
    \centering
    \includegraphics[width=0.7\linewidth]{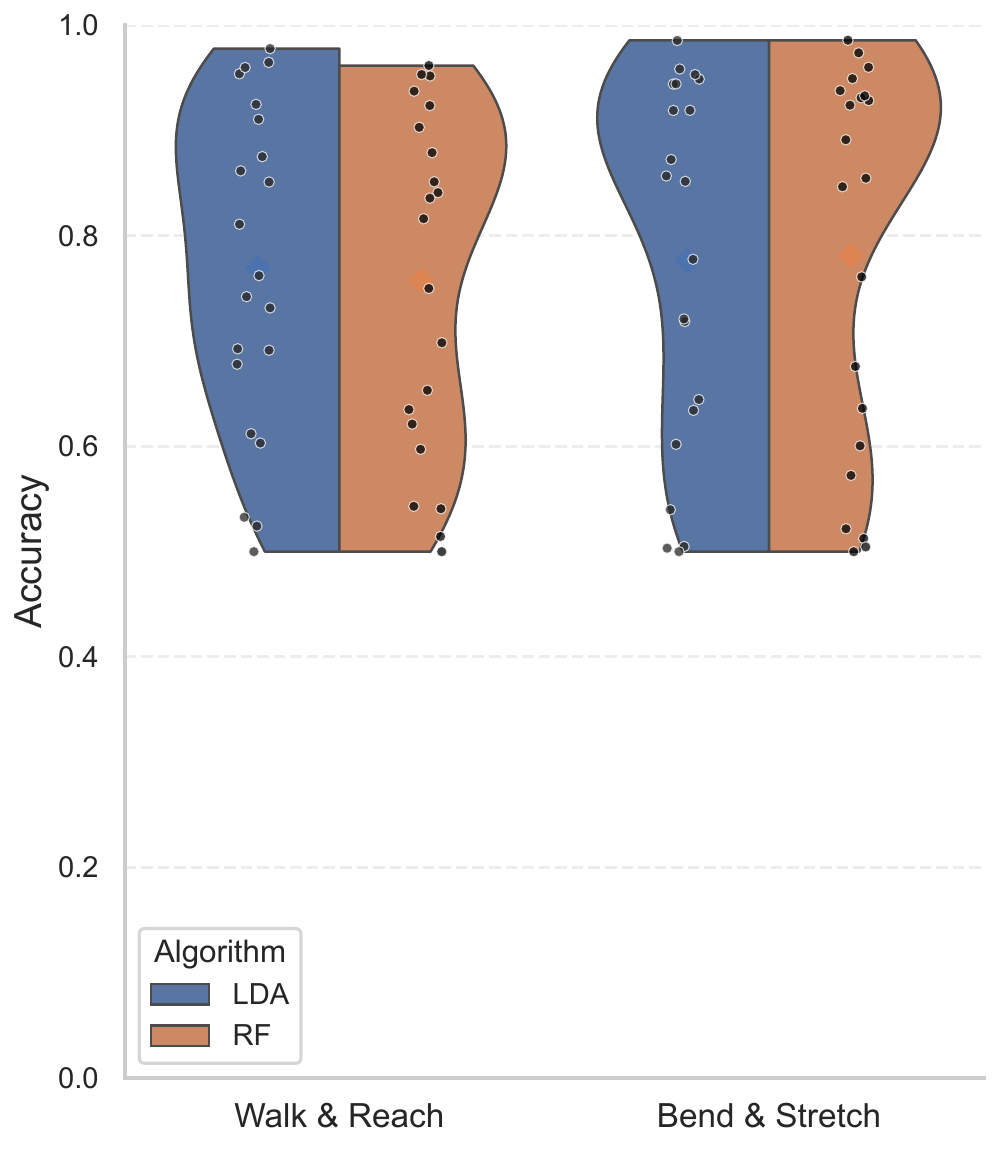}
    \caption{Activity classification by means of the sEMG collected in the MultiSenseVR dataset~\cite{MultisenseVR2026}.}
    \label{fig:sEMG}
\end{figure}
\subsection{Experimental Evaluation of the EMG Signal Processing and Modeling for finger-level interactions} \label{sec:sec:sEMGev}

Remote sensing with 5G has the potential to provide useful information for body pose estimation as shown in the previous results. However, XR interaction also depends on accurate recognition of hand intent during object manipulation and interface control. Many interactive actions rely on subtle differences in grasp type and finger coordination, which are harder to recover from wireless sensing alone because of limited spatial resolution, occlusion, and the small scale of hand motion. As a result, body-level sensing does not fully capture the detail needed for rich controller-free interaction. In this context, EMG is a relevant complementary modality because it measures the underlying muscle activity that drives hand and finger motion. Rather than observing the hand externally, it captures neuromuscular patterns directly from the forearm, making it well suited for interaction cues that may be difficult to infer from RF signals alone. This makes EMG particularly promising for studying fine-grained hand behavior in XR, where subtle finger actions play an important role in grasping, selecting, and manipulating virtual objects.
 
In our previous work, we collected the dataset MultiSenseVR~\cite{MultisenseVR2026}. It includes recordings from 24 participants interacting with a custom VR fast-food simulation designed to elicit natural full-body movement where muscle activity (sEMG) was recorded using two Mindrove EMG armbands\footnote{https://mindrove.com/armband/}, each equipped with eight sEMG electrodes and an integrated IMU. 
To validate the quality of the EMG data, we performed gesture classification experiments on pinch and fist grasps, two common interaction primitives in VR. These gestures were recorded across three experimental conditions: Tutorial, Walk \& Reach, and Bend \& Stretch. The Tutorial condition was used for training, while the ambulatory Walk \& Reach and Bend \& Stretch conditions were used to evaluate how well the learned models generalize under increasingly realistic movement settings with higher levels of motion-related noise. In our baseline evaluation on data from 24 participants, classical models already showed that \gls{sEMG} carries strong discriminative information. After pre-processing and feature extraction, Linear Discriminant Analysis and Random Forest achieved accuracies around 75--80\%, with Random Forest reaching up to 0.78 average accuracy for pinch-versus-fist classification (cf~\cite{MultisenseVR2026}). As shown in Fig~\ref{fig:sEMG}, performance remains promising across both movement settings. While this does not yet form a complete benchmark for fine-grained hand behavior, it shows that the dataset captures gesture-specific neuromuscular patterns relevant to more subtle hand and finger actions. In particular, pinch already represents a more precise hand configuration than gross whole-arm movement and requires coordinated finger activation that is difficult to infer from RF sensing alone.

\section{Conclusion}
In this work, we presented a multi-modal 5G ISAC architecture for immersive XR applications. We showed that low-overhead PPBP can serve as a reliable sensing feature for in-domain gesture recognition, achieving 95.4\% accuracy across 8 users. Further, under the LOUO setting, the PPBP achieves 82.2\% overall accuracy, with 16.4 pp spread across users. We found that LOUO performance can be further improved using feature engineering in the PPBP domain by considering temporal frame differences. 
Finally, lightweight sEMG sensors can augment 5G gesture recognition by identifying fine-grained finger-level interactions. Combining the two modalities will enable multi-scale gesture recognition forming a complete XR framework.
\section*{Acknowledgment}
This research is funded by the Research Foundation Flanders (FWO)
WaveVR project (G034322N) and partly by the FWO SENTIENCE
project (G0A8N25N). Nabeel Bhat is funded by an FWO SB fellowship (1SH5X24N).

\bibliographystyle{IEEEtran}
\bibliography{references}
\end{document}